# How do you know your spreadsheet is right?
## Principles, Techniques and Practice of Spreadsheet Style
### Philip L. Bewig — July 28, 2005

You know it's true: Spreadsheets have errors like dogs have fleas.[1] It is generally accepted[2] that nine out of every ten spreadsheets suffer some error, and consequences can be severe:[3]

- A cut-and-paste error cost TransAlta $24 million when it underbid an electricity-supply contract.[4]

- A missing minus sign caused Fidelity's Magellan Fund to overstate projected earnings by $2.6 billion (yes, *billion*) and miss a promised dividend.[5]

- Falsely-linked spreadsheets permitted fraud totaling $700 million at the Allied Irish Bank.[6]

- Voting officials reported spreadsheet irregularities in New Mexico[7] and South Africa.[8]

- A new drug introduction was delayed several months by an untested macro, costing the pharmaceutical company profits and its patients misery.[9]

You can't eliminate errors from the spreadsheets you develop, but you can reduce their number. The principles and techniques described below, applied consistently, will improve the quality of your spreadsheets. The discussion assumes Excel, but the principles and techniques apply everywhere. The spreadsheet shown below will be used as a practical example:

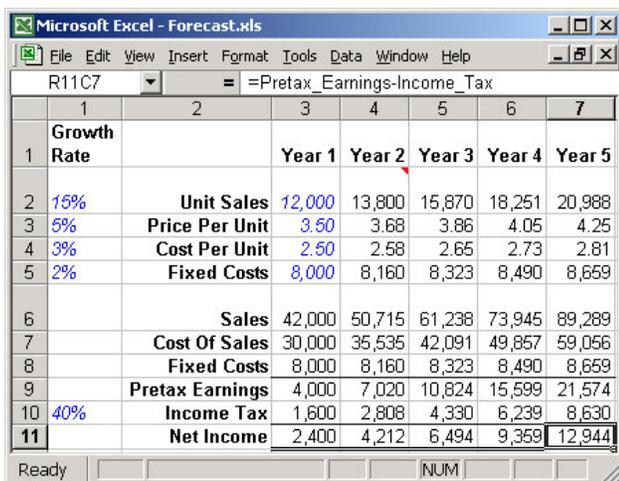

**Think before you write.** Resist the urge to jump right in to actual development. Start with a clear understanding of the requirements of your task. Sketch your design on a whiteboard, and look for flaws. Consider alternate software tools such as databases,[10] statistics packages,[11] financial modeling systems,[12,13] business intelligence systems,[14,15] mathematical programming languages,[16,17,18] and traditional computer programming languages. This is the most fundamental level of your work, and the most creative moment in the entire existence of your spreadsheet. An error here can be hard to fix, requiring massive rearrangements of the spreadsheet structure or new inputs from new sources.

**Know the players.** The *reader* sees the printed output, and uses it to make a decision; he relies on you to organize and present the data he needs, as he needs it. The *user* inputs data, operates macros, and prints output, but doesn't modify anything; he relies on you to provide adequate instructions. You, the *developer*, design and implement the structure and all the formulas in the spreadsheet. The *auditor* checks the work of the developer; he relies on you to produce a clean design and good documentation. The *sponsor* assigns the task, provides resources, and has overall responsibility for the spreadsheet; he relies on you to meet his specifications. In many cases some of these roles overlap; keep them all in mind as you develop your spreadsheet.

**Make your spreadsheet as simple as possible, but no simpler.**[19] Most spreadsheets work well enough with a few SUMs and IFs, and using functions like SUMPRODUCT or features like array formulas, or writing your own macros and functions, can make a spreadsheet harder to read and understand than it should be. On the other hand, don't "dumb down" your spreadsheet, feel free to hide complex logic in user-defined functions, and if some advanced feature will simplify your task, use it. Be wary of features just added to or changed in the newest version of Excel, where bugs are likeliest to lurk; for instance, the RAND function has changed with each version of Excel, and statisticians claim[20,21] it's still not right.

**Plan to throw one away; you will, anyway.[22]** Prototypes are useful for spreadsheet developers for the same reason that scale models are useful for architects; they help you visualize what you are building. They can help you add meaning to ill-defined specifications, demonstrate a partial solution, or work out tricky formulas. Frequently, prototypes grow into a solution; sometimes, the prototype *is* the solution, and the whole problem need never be solved.

**Design for change.** Few spreadsheets are stillborn; most evolve and grow through countless versions, then may be copied with a new name next month when the process starts anew. Brooks says: "All successful software gets changed."[23] The best place to plan for change is during the initial design of the spreadsheet. Reflection, described below, is a useful tool for implementing that plan. Building a change-tolerant spreadsheet isn't much harder than building the other kind, but so much better for the poor fellow who has to modify your work; you'll especially appreciate the initial effort if you are, yourself, that poor fellow.

**Keep input, logic, and reports in separate sections of a spreadsheet, preferably on different tabs.[24]** That way you can always see the assumptions neatly in a single place, formulas are less likely to be overwritten, you know where to go to make changes, and output can be formatted for the reader while logic can be laid out for the developer. If you can't put the three sections on separate tabs, arrange them on a single tab in a stepped diagonal so that rows and columns can be inserted or deleted in each area without affecting the other areas. But beware the false modularity of separate worksheets; since all cells in a worksheet are globally readable, and globally writeable with a macro, using separate worksheets hides no information.[25] Worksheets can't be "dropped in" and reused, nor can they be checked individually without reference to other worksheets.

**Keep your entire spreadsheet on a single tab, intermixing input, logic and reports.[26]** You can't see your whole spreadsheet at a glance if it occupies multiple worksheets. Multiple sheets make formulas longer and harder to read because the sheet name must be included. They breed spurious cells (cells that simply copy other cells without calculation, like =R12C4) because the spreadsheet developer wants to see the precedent cell. The auditing toolbar fails with multiple sheets because arrows don't go to off-sheet cells, and searches are confined to the selected worksheet. With input and logic intermixed, arcs of precedence are shortened.

**Lay out your spreadsheet as determined by the needs of your problem.** Obviously, the two previous suggestions conflict, and in fact no single design is always right; the size of the spreadsheet, frequency of off-sheet references, complexity of formatting, and many other factors must all be considered. The sample spreadsheet has four sections: growth rates (input) at the top of the first column, tax rate (input) in an unlabelled cell at the bottom of the first column, base values (input) and growth amounts (calculated) in the top rows, and income statement (calculated) in the bottom rows. An alternate layout would have base values in a section next to growth rates, with formulas that copy base values into the calculation area of the spreadsheet; this design works well if there are many input values or the calculation section is large. Some people will object to the inclusion of fixed costs twice on the spreadsheet, saying that one or the other is spurious and should be removed, but that's a consequence of separating logic from output; in some cases it will be sufficient to have only a single appearance of the number, but if logic and output are both large, it may make sense to have the number appear twice.

**Make your spreadsheet read top-to-bottom and left-to-right.** All dependent arrows should point down, right, or somewhere in between.

One exception is when the beginning balance at the top of one column depends on the ending balance at the bottom of the previous column.

**Build a complicated spreadsheet in stages.** Let it grow, but always with a working partial solution at hand. Test as you go,[27] so you have confidence in the pieces as well as the whole. Fix problems immediately; don't leave them for the next version.

**Draw the dependency graph.** "A picture is worth a thousand words."[28] The dependency graph of the sample spreadsheet looks like this:

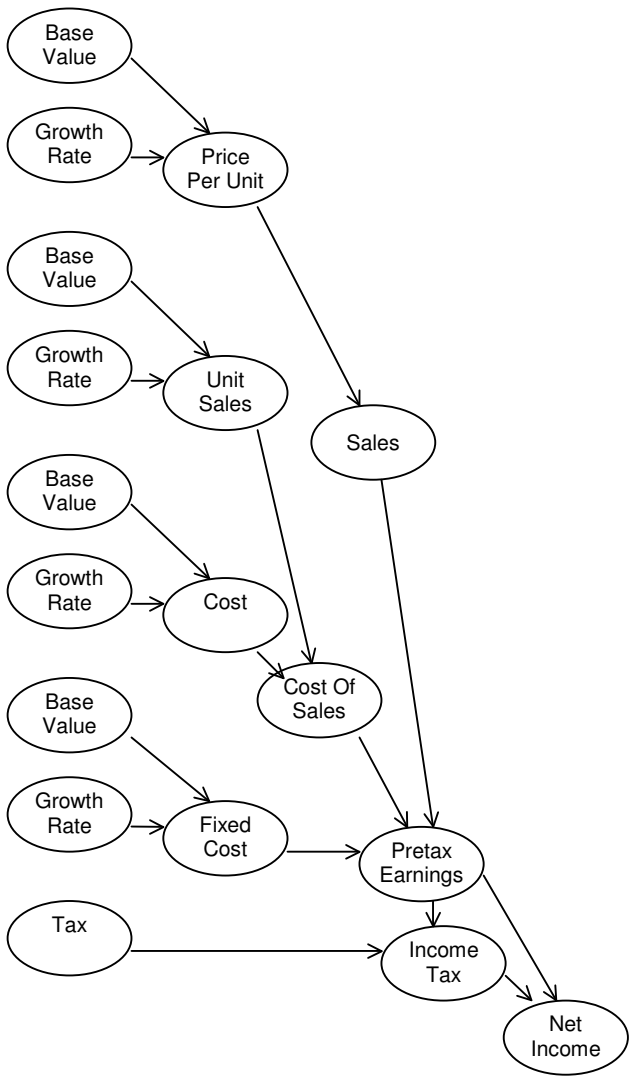

**Beware the cascade effect.** The likelihood that a spreadsheet produces erroneous output is a function of the error rate $e$ for individual cells and the number $n$ of cells that must each be correct in succession (a "cascade" of cells) in order for the whole spreadsheet to be correct. Mathematically,[29] this function is $1-(1-e)^n$, which grows asymptotically to 100%, as shown in the graph below; with a 5% error rate, even a cascade of only six cells gives more than a 25% chance of overall error. You can reduce the cell error rate by careful checking, but it's generally easier to restructure the computation to reduce the length of the cascade. The sample spreadsheet has seventeen cascades, six of six cells, eight of five cells, two of four cells, and one of three cells, as shown in the diagram above (every cascade through Pretax Earnings counts twice, since it has two out-arrows); it also has a one-cell cascade for the year captions that is not included in the diagram.

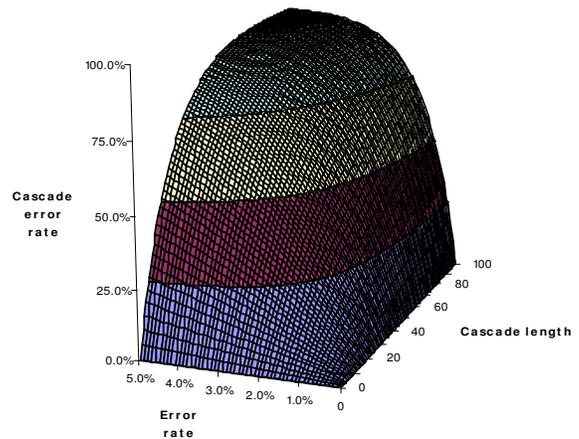

**Document the design of your spreadsheet on a separate HOWTO tab.** State the purpose of your spreadsheet in a single sentence. Include a drawing of the dependency graph. Make a list of all the individual tabs in your spreadsheet, and write a single sentence describing each one. State the source of all inputs (be specific: "Pat at extension 3220 in the Sales Department"), and specify by name and job title all the people that will see your output. Briefly describe your overall design (if that takes more than a sentence or two, your design is too complicated) and point out any unusual or tricky spots. Describe macros and user-defined functions. Include instructions so someone else can change input and obtain output without your help. Update your documentation when the spreadsheet changes.

**Provide basic documentation with `Workbook Properties`.** And make it easily accessible by setting Prompt for workbook properties. It's a convenient place to put summary documentation, and is searchable. The Category and Keyword fields are useful (think about a category `July2005` for all your workpapers this month), and the Custom tab provides many additional fields, including `Purpose` and `Checked by`.

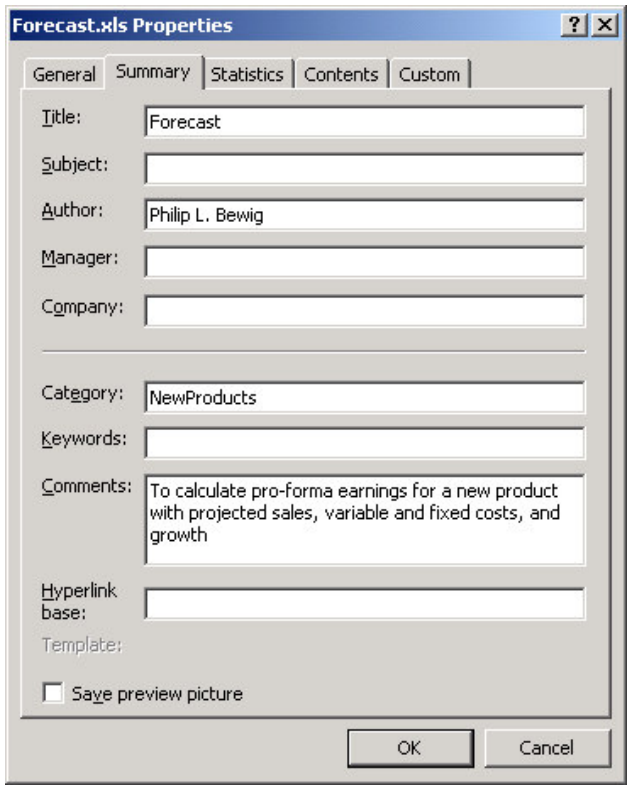

**Use `R1C1`-style cell references.** Though `$A$1`-style cell references are more common, `R1C1`-style cell references are preferable because they are self-contained; you don't need to know the current address to know what relative references in a formula mean. For instance, `=C7+C8` means something different in cell `C9` than in cell `C10`, but `=R[-2]C+R[-1]C` means the same thing no matter the current cell, and copies are visually identical to the original.

**Use descriptive range names.** But don't use Excel's natural-language labels, which sometimes fail in unexpected ways. It's easier to read `=Sales-Cost_Of_Sales-Fixed_Cost` than `=R[-3]C-R[-2]C-R[-4]C` and understand that it says what you expect.[30] Cell addresses are a physical concept; names are a higher-level logical concept directly relating spreadsheet to task. Names should be meaningful, brief, and distinctive;[31] well-chosen names are the first and best form of documentation. Names are better declared with `Create` rather than `Define`, as there is less possibility of the name referring to the wrong range. Prefer VBA functions to named formulas. Here is a list of names defined in the sample spreadsheet, produced by Paste List:

```
Cost_Of_Sales      =Forecast!R7C3:R7C7
Cost_Per_Unit      =Forecast!R4C3:R4C7
Fixed_Costs        =Forecast!R5C3:R5C7
Growth_Rate        =Forecast!R2C1:R5C1
Income_Tax         =Forecast!R10C3:R10C7
Pretax_Earnings    =Forecast!R9C3:R9C7
Price_Per_Unit     =Forecast!R3C3:R3C7
Prior_Year         =Forecast!RC[-1]
Sales              =Forecast!R6C3:R6C7
Tax_Rate           =Forecast!R10C1
Unit_Sales         =Forecast!R2C3:R2C7
Year               =Forecast!R1C3:R1C7
```

**Thoroughly understand the difference between absolute and relative references.** Like pointers in a traditional programming language, errors in absolute and relative references can cause insidious errors that are almost impossible to find. Be sure you understand the different effects of insertion, deletion, copy, sorting, and other ways of moving a cell from one place to another.

**Know the difference between early-binding and late-binding cell references.** Normal cell references, both absolute and relative, using either $A$1-style or R1C1-style cell references, are early-binding; if you insert a row in a column of formulas, each referring to the one above, the cell below the insertion point will continue to refer to the cell that was originally above it even after the insertion. Range names and cells referenced by `OFFSET` or `INDIRECT` are late-binding; if that same insertion was done using a range name that referred relatively to the cell above, the cell below the insertion point would refer to the new cell.

**Allow only one unique formula per row or column.** Consider your spreadsheet as a database table, with attributes (fields) and tuples (records); for instance, the sample spreadsheet has income and expenditure captions running down the leftmost column and time marching along the top row, giving it row attributes and column tuples. One useful design technique is to organize your spreadsheet so each attribute uses only a single formula. Thus, you should prefer

|  | This year | Last year |
|---|---|---|
| Sales | $14,729 | $14,021 |
| Gross profit | $4,601 | $4,292 |
| Net income | $1,245 | $880 |
| Percent to sales |  |  |
| Gross profit | 31.2% | 30.6% |
| Net income | 8.5% | 6.3% |

rather than

|  | This year |  | Last year |  |
|---|---|---|---|---|
| Sales | $14,729 |  | $14,021 |  |
| Gross profit | $4,601 | 31.2% | $4,292 | 30.6% |
| Net income | $1,245 | 8.5% | $880 | 6.3% |

Your reader will prefer it, too, since the various elements of the analysis are more clearly separated. If a single attribute must use two formulas, write them in the two legs of an IF, using a reflective condition to distinguish them; for instance, the formula =(1+IF(Year<=3,Growth_Rate,MIN(10%,Growth_Rate))*Prior_Year caps the growth rate at 10% after the third year.

**Consider entering identical formulas in adjacent cells as array formulas rather than copies.** Since all cells in the array are identical, by definition, it is impossible to sustain the very common spreadsheet error of changing some but not all of a series of copied formulas. But array formulas[32] can't be recommended in most cases, since they freeze the structure of a spreadsheet, rendering subsequent changes inconvenient. The sample spreadsheet uses copies rather than array formulas.

**Make assumptions visible, but hide magic numbers.** Use numbers explicitly in formulas only when they represent mathematical identities, as in (1+GrowthRate)*PriorYear. Magic numbers (constants that are an artifact of the implementation rather than the problem, such as the offset in a table lookup) are best defined as named constants, where they are visible to the spreadsheet developer (and easy to maintain) but hidden from the reader. Assumptions (numbers intrinsic to the problem, such as tax rates) should be made visible in their own cells.

**Validate input cells, and make them visible with a "glowing pencil."[33]** Use the BETWEEN operator, make ranges tight enough to be effective, and never use open-ended ranges; enumerated lists are also effective. Use the input prompt to instruct the user (even if that user is yourself) of the source of the input. Turn off auto-completion; it's easy to automatically insert incorrect data. Ensure both you and your reader know which cells are input cells by formatting them differently than the rest; blue italics work well, distinguishing cells on-screen and when printed. Here is the cell validation list for the sample spreadsheet:

```
R2C1     Decimal        Between   0%    100%
R3C1     Decimal        Between   0%    100%
R4C1     Decimal        Between   0%    100%
R5C1     Decimal        Between   0%    100%
R10C1    Decimal        Between   0%    100%
R2C3     Whole Number   Between   0     100,000
R3C3     Decimal        Between   0     1,000
R4C3     Decimal        Between   0     1,000
R5C3     Whole Number   Between   0     100,000
```

That list was created by the following macro:

```
Sub PasteValidationTable()
    Dim C As Range, T As Range
    Set T = ActiveCell
    For Each C In ActiveSheet.Cells.Special-
            Cells(xlCellTypeAllValidation)
        T.Offset(0, 0) = _
            C.AddressLocal(True, True, _
            Application.ReferenceStyle)
        T.Offset(0, 1) = _
            Choose(C.Validation.Type + 1, _
            "Input Only", "Whole Number", _
            "Decimal", "List", "Date", _
            "Time", "Text Length", "Custom")
        T.Offset(0, 2) = _
            Choose(C.Validation.Operator, _
            "Between", "Not Between", "Equal", _
            "Not Equal", "Greater", "Less", _
            "Greater Or Equal", "Less Or Equal")
        T.Offset(0, 3) = C.Validation.Formula1
        T.Offset(0, 4) = C.Validation.Formula2
        Set T = T.Offset(1, 0)
    Next C
End Sub
```

**Be wary of links to other spreadsheets or external data sources.** A changed link that auditors failed to catch was the trick behind the Allied Irish Bank fraud; the criminal spreadsheeter substituted a spreadsheet containing his made-up cross-currency rates for one maintained by the bank. It's hard to know where links point; they tie the spreadsheet to a particular directory structure, and may mask circular references. Links expand audits; link-ees must be audited as well as link-ers. And links lose synchronization if not updated. Format linking cells distinctively; violet italics work well.

**Trap formula errors.** But allow unexpected errors to propagate rather than masking them. Convert errors into appropriate values; `ISERROR` and `ISNA` may be helpful. Any reader who sees `#DIV/0!` is entitled to assume your incompetence without further evidence.

**Maintain a common interval for calculations.** If your spreadsheet switches from quarterly forecasts for the first year to annual forecasts thereafter, keep calculations in quarters, then sum the out-year quarters for reporting. Alternately, write two separate spreadsheets.

**Describe your formulas.** If they are complicated or there is some possibility of confusion, describe your formulas, in English prose, directly on the printed output. That way, both you and your reader know what you are doing.

**Use `Goal Seek`[34] to calculate iterative solutions.** Reserve circular references to indicate errors. Use the circular reference toolbar to track down circular references that do occur.

**Lock all cells except input cells, but publish the key.** Your goal is to prevent inadvertent change, not to forestall needed change. Excel's password protection is easy to defeat,[35] so don't rely on it for security. Be sure to know all your options, including file-level security from the operating system, and the various levels of protection that Excel offers to workbooks, worksheets, and cells.

**Program defensively using assertions.** Given

```
Function Assert(X, Y, Msg As String)
    If Abs(X / Y - 1) < 1E-13 Then
        Assert = X
    Else
        Assert = 1 + Msg
    End If
End Function
```

then `=ASSERT(SUM(RowSums), SUM(ColSums), "Row sums must equal column sums")` in the lower right-hand corner of a table will check that the table foots and cross-foots; the strange-looking condition traps rounding differences. If the assertion ever fails, the expression `1+Msg` will throw an error that propagates to the final result; backtracking to the source of the error will bring you to the message planted earlier. Assertions can be based on mathematical identities (footings and cross-footings must equal), external identities (assets equal liabilities plus equities), or redundancies (two different inputs both lead to the same output). Assertions[36] are valuable because they test the spreadsheet dynamically, every time it is recalculated, instead of statically when a test is called. If you wish, you could replace the growth formulas in the top half of the sample spreadsheet with the assertion suggested in the comment box below.

**Document your work liberally using in-cell comments.** Describe precisely the source of all inputs. Explain complicated formulas. Document user-defined functions and macros. Explain the condition and both arms of any `IF`s. Document your work as you are doing it, when the reason for all your design decisions is fresh in your mind; you will surely forget something important if you save the documentation work for later. Update comments when the cell contents change. This comment is in cell R2C4 of the sample spreadsheet:

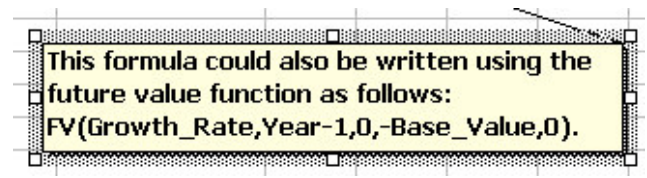

This formula could also be written using the future value function as follows: FV(Growth_Rate,Year-1,0,-Base_Value,0).

**Use Excel's reflective features to automate your work.** Reflection[37] makes a programming language *self-referent*. The `CELL`, `INFO` and `TYPE` functions return much useful information. Dynamic named ranges[38] grow and shrink with the data. Use `COUNTA` instead of hard-wiring the size of a range. Define names as entire rows or columns instead of just occupied cells. Write macros that adapt to the size of the data instead of fixing the number of cells in advance. Get familiar with the tasteful use of the `OFFSET` and `INDIRECT` functions. When you use reflection, the computer does more of the work and you do less of it; guess which of you is more likely to get the right answer?

**Build reflective features into your spreadsheets.** The sample spreadsheet uses the formula `=1+Prior_Year` to put the numbers 1 through 5 in the year-caption cells, displays the cells with the custom format `"Year "#`, and assigns the range name `Year`. That makes it easy to refer to the year number, as in the previous example capping the growth rate.

**Exploit the `Go To Special` dialog.** It selects special cells on the active worksheet; for instance, click Constants and Numbers to select all cells containing a number, or click Formulas and Errors to select all cells containing error values like `#N/A!` or `#DIV/0!`. The Row differences and Column differences options are especially useful in discovering formulas that differ from their neighbors.

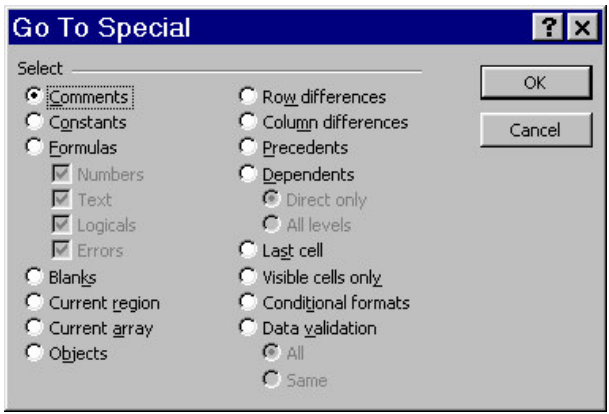

**Use conditional formatting to highlight unlikely results.** Or collect all your sanity checks at the end of the spreadsheet. If the new product forecast on the sample spreadsheet suggests a 57% net income, it's either a great new product or there's some error in model or data.

**Format using styles rather than toolbar buttons.** In fact, you should close the formatting toolbar entirely. Styles can be applied quickly once they are set up and provide a consistent interface to the reader; you may want to add the style control to the standard toolbar. Save common styles in a template in your XLStart folder. Even better, program common styles on a toolbar loaded by an add-in, and have everybody in your department use the same formatting styles, giving a consistent look everywhere.

**Use blanks around rows and columns to define "areas."** Areas (a maximal rectangle of occupied cells surrounded by unoccupied cells) have meaning to Excel's object model, define the extent of an `END+ARROW` key, and can be unioned to form a range. It may be useful to make each distinct area a separate worksheet. Instead of blank rows and columns, create output space using row height and column width, and use a single cell with embedded newlines to store a lengthy column header.

**Avoid color.** Although some printers now handle color, most photocopy and fax machines still don't, so assume a black-and-white world. That's limiting only if you let it be; remember Apple's original Macintosh was gray-scale, but the crisp, clear screen made many color screens look blurry and ugly by comparison. Also avoid shaded backgrounds that lose contrast.

**Don't let your charts look like ducks.[39]** Line charts present continuous data, bar charts present discrete data, and pie charts show the components of a single variable. Area charts and stacked bar charts show how the components of a variable fluctuate over time, for continuous and discrete variables, respectively. A scatter chart shows the relationship between two discrete variables. Charts with a false third dimension should be avoided, since the extra dimension adds clutter without adding meaning.

**Never do the same thing twice; record a macro instead.** The macro will do the job much faster than you can, and more consistently. It's not hard[40] to write simple macros, especially if you use the macro recorder, and the effort you put into learning to write macros will be amply repaid in improved accuracy and reduced development times.

**Break up a complicated expression into multiple cells with intermediate results.** The result is generally more readable than one big formula could ever be. Often, the intermediate results are useful, too. If you wish, ASSERT that the final answer is the same as the complicated expression. But consider the earlier advice about cascade lengths, and consider the next technique, which provides an alternative.

**Write a VBA function rather than a complicated expression.** VBA provides more built-in power than Excel expressions, so VBA functions can do more things more easily. If you can write Excel expressions you can write functions, at least simple functions that avoid such features as loops and file operations.

**Let the VBA development environment help you.** The single most important rule is to automatically set OPTION EXPLICIT by clicking on Require Variable Declaration; think of it as a prophylactic for your code. Make friends with the Object Browser, and learn to find objects in the help screens. Use the debugger to set watch variables and breakpoints and single-step your code.

**Build a library of frequently-used macros and functions.** Store the library in your PERSONAL.XLS spreadsheet so it is always available. But be aware that your personal macros and functions won't be available if you share your spreadsheet with someone else. If you prefer, common macros and functions can be distributed in an add-in.

**Build templates for common tasks.** Templates store captions, formulas, formatting, and macros, making it easy to build a set of similar spreadsheets. Templates speed debugging as well as development because the common portions only need to be debugged once.

**Use a macro to automatically add the user name, file name, date and time to the footer of all your spreadsheets.** Your readers deserve to know. A macro stored in an add-in that traps the `WorkbookBeforePrint` event is simple and convenient, and can use VBA functions to extract `Workbook Properties` and perform other useful tasks. You may want to have the same macro force a file-save with a new name, so you can quickly return to any printed version of your spreadsheet.

**Print a listing of all your formulas.** If you follow the earlier advice to name all cells and write one formula per row or column, you can print a list of all your formulas by transposing the spreadsheet so attributes run down the left column, displaying formulas rather than values, and printing. The resulting listing shows all the logic of your spreadsheet, neatly formatted.[41] It's much easier to read the printed listing than to scroll back and forth from cell to cell on the screen; when you look at cell addresses on screen, you tend to think physically ("that cell over there") rather than logically ("unit sales"), and what do you do if "over there" has scrolled off the screen? That listing, with all your hand-written notes and marks, is an excellent exhibit to show the auditors why you think your spreadsheet is right; it's even useful to show your boss how hard you worked. A formula listing of columns C2 and C4 of the sample spreadsheet is shown below:

```
Year            =1+Prior_Year

Unit_Sales      =(1+Growth_Rate)*Prior_Year
Price_Per_Unit  =(1+Growth_Rate)*Prior_Year
Cost_Per_Unit   =(1+Growth_Rate)*Prior_Year
Fixed_Costs     =(1+Growth_Rate)*Prior_Year

Sales           =Unit_Sales*Price_Per_Unit
Cost_Of_Sales   =Unit_Sales*Cost_Per_Unit
Fixed_Costs     =Fixed_Costs
Pretax_Earnings =Sales-Cost_Of_Sales-Fixed_Costs
Income_Tax      =Pretax_Earnings*Tax_Rate
Net Income      =Pretax_Earnings-Income_Tax
```

**Use a spreadsheet auditing tool.** Excel's auditing toolbar allows you to trace precedents and dependents and highlight invalid data, but is otherwise limited. Spreadsheet auditing tools[42] help you identify errors with a variety of identification, searching, reporting, and visualization techniques; using them will help you better understand your spreadsheet. My CellMaps[43] tool colors cells according to type, as in this display of the sample spreadsheet, showing input cells with pink backgrounds, formulas with lavender backgrounds, and formulas copied right, down and both with blue, green, and gray backgrounds, respectively.

[Screenshot of Microsoft Excel - Forecast.xls showing a spreadsheet with R1C1-style references. Selected cell R6C7 contains =Unit_Sales*Price_Per_Unit. Table contents:

| | 1 | 2 | 3 | 4 | 5 | 6 | 7 |
|---|---|---|---|---|---|---|---|
| | Growth Rate | | Year 1 | Year 2 | Year 3 | Year 4 | Year 5 |
| 2 | 15% | Unit Sales | 12,000 | 13,800 | 15,870 | 18,251 | 20,988 |
| 3 | 5% | Price Per Unit | 3.50 | 3.68 | 3.86 | 4.05 | 4.25 |
| 4 | 3% | Cost Per Unit | 2.50 | 2.58 | 2.65 | 2.73 | 2.81 |
| 5 | 2% | Fixed Costs | 8,000 | 8,160 | 8,323 | 8,490 | 8,659 |
| 6 | | Sales | 42,000 | 50,715 | 61,238 | 73,945 | 89,289 |
| 7 | | Cost Of Sales | 30,000 | 35,535 | 42,091 | 49,857 | 59,056 |
| 8 | | Fixed Costs | 8,000 | 8,160 | 8,323 | 8,490 | 8,659 |
| 9 | | Pretax Earnings | 4,000 | 7,020 | 10,824 | 15,599 | 21,574 |
| 10 | 40% | Income Tax | 1,600 | 2,808 | 4,330 | 6,239 | 8,630 |
| 11 | | Net Income | 2,400 | 4,212 | 6,494 | 9,359 | 12,944 |
]

**Avoid common errors.** The single most common error[44] in spreadsheet development is pointing to the wrong cell during formula construction; fortunately, use of named ranges eliminates that problem, as long as you construct the range names automatically. Other common errors include:

- Changing some but not all of a series of copied cells (array formulas prevent this problem, but at considerable inconvenience);
- Incomplete ranges (cells are omitted from one end of a range or the other, often when a range is expanded but the formula that relies on the range isn't changed; this was the error in the TransAlta case);
- Temporary fixes (a calculated cell is replaced by the amount it calculates, perhaps to make the rounding come out right, but is not changed back to a formula when the input values change later);
- Confusion between relative and absolute references (R1C1-style cell references clarify the difference, and named ranges help, but training is the only answer for this problem);
- Incorrect units (mixing apples and oranges, this problem is frequently seen when mixing units with thousands of units); and
- Function arguments in the wrong order (this error can be hard to find, but using the function wizard to help write unfamiliar functions can make it less frequent).

Even more common than these technical errors is failing to meet the specifications of the task.[45] It never hurts to write down exactly what you intend to do, and get approval from your sponsor before you proceed.

**Check your work before you publish it.** Calculate check totals. Compare to previous work. Test with known values. Put zeros in all the input cells, and ensure the output is zero. Perform dimensional analysis,[46] to make sure the units come out right. Perform sensitivity analysis; does adding one more unit sold increase pretax earnings by the gross profit of one unit? Stress test your spreadsheet at the outer limits of all data validation bounds. Sit down and read a printed version of your spreadsheet, as if you were the intended reader. Approach your spreadsheet as if you were an auditor examining any financial exhibit; test the footings with a calculator and tie all inputs to source documents. Wait until morning, then check it again. Best is to document your test plan before you begin work, then perform your test plan before you publish. Spreadsheets that are large, complex, of great import or urgency, or that have ill-defined specifications bear more risk and thus require more testing than spreadsheets without those properties. Don't fret about errors your tests reveal; remember the purpose of testing is to find the errors that certainly exist, so finding an error is a vindication of your test work, not an indictment of your development work.

**Have a colleague check your work before you publish it.** He should review your documentation, analyze your design, critique your test plan, and test for reasonableness. Cell-by-cell inspection is tedious, but is the only method of spreadsheet testing known to be effective.[47] My TrafficLights[48,49] tool, shown in action below, provides a simple framework for cell-by-cell inspection. The auditor has already checked green cells, finding no errors. Yellow cells are ready to check because, as the arrows show, all their precedents are green. The auditor is about to check R7C3; adjacent cells with the same formula are a darker yellow, suggesting to the auditor that he may want to check them at the same time. Red cells are not ready to check because they have unchecked precedents. When the entire spreadsheet is green, the audit is complete.

**Work in pairs.** Panko and Halvorsen[50] report that the effect of combining the efforts of multiple spreadsheet developers working together in front of a single computer is positive, reducing errors as each checks the work of the other. One approach to pair programming[51] is to couple a spreadsheet expert with a domain expert; their skills complement each other. Another approach is to couple a domain expert with someone junior (or someone expert in a different domain), to allow skills transfer between the two. Two or more auditors also seem to be better than one; related work by the same authors shows[52] that two auditors who check independently then compare notes will find over four-fifths of all errors, as compared to the third-to-a-half of errors found by a single auditor.

**Create an organizational environment that encourages correctness.** Insist on printing the developer's name on each page of a spreadsheet; asking people to acknowledge their work is a powerful incentive for them to get it right. Establish a formal system of peer review of all spreadsheets; be sure to add the reviewer's name to the identification string in the footer. Print and store documentation and formula listings in a standard location and periodically pick one for review and group discussion. For critical spreadsheets, two developers can work independently writing two separate spreadsheets, then compare, cycling until both agree.

**Maintain a log of spreadsheets in use.** Review the log with your boss monthly. Mark critical spreadsheets for audit, and allocate the needed time. Don't maintain your log by hand; instead, write a macro that crawls your file tree and extracts the documentation from `Workbook Properties`. Called as `PasteSpreadsheetLog "C:\", ActiveCell`, the crawler shown below is a useful starting point for a crawler specific to your needs:

```
Sub PasteSpreadsheetLog( _
  ByVal DirName As String, _
  OutCell As Range, _
  Optional Headers As Boolean = True)

  Dim CurrFile As String, CurrDir As String
  Dim DirNames() As String, DirIndex As Integer
  Dim InBook As Workbook, OutBook As Workbook

  DirIndex = 0
  Set OutBook = ActiveWorkbook
  Application.ScreenUpdating = False

  If Right(DirName, 1) = "\" Then
    DirName = Left(DirName, Len(DirName) - 1)
  End If

  If Headers Then
    OutCell.Offset(0, 0) = "Directory Name"
    OutCell.Offset(0, 1) = "File Name"
    OutCell.Offset(0, 2) = "File Size"
    OutCell.Offset(0, 3) = "Author"
    OutCell.Offset(0, 4) = "Comments"
    OutCell.Offset(0, 5) = "Checked By"
    OutCell.Offset(0, 6) = "Purpose"
    Set OutCell = OutCell.Offset(1, 0)
  End If

  CurrFile = Dir(DirName & "\*.xls", vbNormal)
  Do Until CurrFile = ""
    If CurrFile <> OutBook.Name Then
      Workbooks.Open _
        FileName := DirName & "\" & CurrFile
      Set InBook = ActiveWorkbook
      On Error Resume Next
      OutCell.Offset(0, 0) = DirName
      OutCell.Offset(0, 1) = CurrFile
      OutCell.Offset(0, 2) = _
        FileLen(DirName & "\" & CurrFile)
      OutCell.Offset(0, 3) = InBook.-
        BuiltinDocumentProperties("Author")
      OutCell.Offset(0, 4) = InBook.-
        BuiltinDocumentProperties("Comments")
      OutCell.Offset(0, 5) = InBook.-
        CustomDocumentProperties("Checked by")
      OutCell.Offset(0, 6) = InBook.-
        CustomDocumentProperties("Purpose")
      On Error Goto 0
      InBook.Close SaveChanges := False
      Set OutCell = OutCell.Offset(1, 0)
      CurrFile = Dir
    End If
  Loop
```

```
  CurrDir = Dir(DirName & "\", vbDirectory)
  Do Until CurrDir = ""
    If CurrDir <> "." And CurrDir <> ".." Then
      If (GetAttr(DirName & "\" & CurrDir) _
          And vbDirectory) = vbDirectory Then
        DirIndex = DirIndex + 1
        ReDim Preserve DirNames(DirIndex)
        DirNames(DirIndex) = _
          DirName & "\" & CurrDir
      End If
    End If
    CurrDir = Dir
  Loop

  Do While DirIndex > 0
    CurrDir = DirNames(DirIndex)
    PasteSpreadsheetLog CurrDir, OutCell, False
    DirIndex = DirIndex - 1
  Loop

  Application.ScreenUpdating = True
End Sub
```

The curious structure of this code is due to the `Dir` function, which is non-reentrant,[53] requiring that directory names be fetched and stored rather than processed recursively as they occur.

**Turn on autosave or autorecover.** There is no excuse for you to lose your work just because your computer crashed. You might also want to checkpoint your work by periodically saving it with a new name, so that if you make an unrecoverable mistake during development, you won't lose too much time. And turn on automatic backups; it's cheap insurance against tragedy. Although autosave, checkpointing and automatic backup are similar, they serve different purposes; do them all.

**Master your tools.** Spreadsheet programs are big, and there is always more to learn, even for experts. Public libraries and community colleges may have training classes available, or you may prefer to work on your own, with one of these sources, among many others: John Walkenbach has written several books about Excel, and maintains a useful website.[54] Microsoft provides a guide to the Excel object model,[55] and much other useful information at its support website.[56] The Journal of Accountancy publishes a Technology Workshop series with frequent articles about Excel.[57] Several Usenet newsgroups discuss Excel and other spreadsheet programs.[58,59] MVPs, the Microsoft Most Valuable Professionals, provide much useful advice, and frequently answer specific questions posted by spreadsheet developers on Usenet.[60] The European Spreadsheet Risks Interest Group hosts a website,[61] a discussion group,[62] and an annual conference.

**Get in the habit of reading spreadsheets.** Writers read books. Filmmakers watch movies. Athletes study slow-motion video of their competitors. Likewise, you should read spreadsheets to learn new techniques, be better able to compare the relative strengths and weaknesses of different techniques, and be more critical when reading your own spreadsheets.

**Be humble.** Panko and Halverson report a study[63] in which developers prior to examination of their spreadsheets expected only an 18% error rate but found an actual 86% error rate. Overconfidence works like this: when you don't look for errors, you don't find them, giving you false confidence in your low error rate; when you later stumble upon an error, the event is rare, reinforcing your notion of a low error rate. The key to breaking overconfidence is to analyze every spreadsheet you see, so you will recognize the signature of errors, and take steps to reduce them in your own work.

**Expect errors.** As a reader, never pick up a spreadsheet and use it without first assuring yourself it is correct; even following all the principles and techniques suggested here won't reduce an average 90% error rate to zero, at least not all the time. Simply relying on someone else's work doesn't absolve you of the effects of any errors that you derive from it.

**Break the rules.** These are principles and techniques, not rules, and you should feel free to ignore any of them, provided that you can articulate how the benefit to be derived from doing so exceeds the cost of the infraction. Writing a spreadsheet is, in many ways, the same type of intellectual activity as writing English prose. Says[64] William Strunk, Jr., in the introduction to his classic English composition textbook *The Elements of Style*:

> It is an old observation that the best writers sometimes disregard the rules of rhetoric. When they do so, however, the reader will usually find in the sentence some compensating merit, attained at the cost of the violation. Unless he is certain of doing as well, he will probably do best to follow the rules. After he has learned, by their guidance, to write plain English adequate for everyday uses, let him look, for the secrets of style, to the study of the masters of literature.

Strunk's advice is a century old, and intended for a different medium, but still speaks today.

---

There have been other attempts to define a standard set of rules of spreadsheet style,[65,66,67] and there is no reason that spreadsheet style can't evolve over time. But even though recent thinking is that no one set of rules will work in all cases,[68] this paper attempts to distill some basic principles and techniques that apply everywhere.

Academic experiment, common sense, painful experience, and crossover from other forms of computer programming are the root sources of the principles and techniques described above. The basic ideas are few: keep it simple, strive for clarity, use tools appropriately, and plan for change, so you won't go too far wrong. Excel provides a rich programming environment, with powerful operators and multiple ways to achieve the same result. The principles and techniques described above seek to exploit the richness while limiting alternatives in ways that prevent errors.

Styling a spreadsheet requires demanding intellectual effort, and if you are adept at building spreadsheets, you may think the principles and techniques described here just slow you down without providing any offsetting benefit. If you feel that way, try this experiment: audit one of your spreadsheets, then, for two weeks, build all your spreadsheets honoring all these principles and techniques, exactly, so you're well through the learning curve, then audit one of your new spreadsheets, and decide for yourself if the quality of your spreadsheets has improved. Once you determine your style, make it a habit, so you focus on the particular needs of your problem rather than the style with which you write it.

It's good to ask "Is your spreadsheet right?" but confidence comes from answering the question "How do you know your spreadsheet is right?" That's a really tough question. If all you can say is that you're a Spreadsheet Super Man, you should hang up your blue tights and red cape and stop writing spreadsheets. But it would be most impressive if you could say

> I followed accepted best-practice in developing my spreadsheet. I designed my work carefully, and reviewed it when I was finished, using a commercial auditing tool. I tested the spreadsheet using known results, according to a written test plan, and compared to other similar work. Knowing that the result was important, I had Bob, Mary and Joe peer-review my work. There could still be mistakes, but I was at great pains to prevent them.

What's your answer?

---

[1] John F. Raffensperger, "New Guidelines for Spreadsheet Style," *Proceedings of the European Symposium on Spreadsheet Risks*, Amsterdam, July, 2001, pages 61–76, www.mang.canterbury.ac.nz/people/jfraffen/spreadsheets/.

[2] Raymond R. Panko, "What We Know About Spreadsheet Errors," *Journal of End User Computing's Special issue on Scaling Up End User Development*, Volume 10, Number 2, Spring 1998, pages 15–21, panko.cba.hawaii.edu/ssr/Mypapers/whatknow.htm.

[3] European Spreadsheet Risks Interest Group, *Spreadsheet Horror Stories*, www.eusprig.org/stories.htm.

[4] Drew Cullen, "Spreadsheet snafu costs firm $24m," *The Register*, June 19, 2003, www.theregister.co.uk/content/67/31298.html.

[5] Kathy Godfrey, "Computing error at Fidelity's Magellan Fund," *The Risks Digest*, Volume 16, Issue 72, January 6, 1995, catless.ncl.ac.uk/Risks/16.72.html.

[6] Ray Butler, "The Role of Spreadsheets in the Allied Irish Banks / Allfirst Currency Trading Fraud," *Proceedings of the 2nd Annual EuSpRIG Symposium*, Amsterdam, Netherlands, July 2001, www.gre.ac.uk/~cd02/eusprig/2001/AIB_Spreadsheets.htm.

[7] Ed Asher, "Glitch scrambles city election tally - computer error changes none of the outcomes," *Albuquerque Tribune*, November 8, 2003, http://web.abqtrib.com/archives/news03/110803_news_canvass.shtml.

This document is available electronically as `HowDoYouKnowYourSpreadsheetIsRight.pdf` in the Files section of the European Spreadsheet Risks Interest Group discussion group at groups.yahoo.com/groups/eusprig. That same web site is an appropriate place for discussion.

Philip L. Bewig, MBA, CPA, first used Multiplan in 1983, and has subsequently used 1–2–3, Excel, and OpenOffice. Phil gratefully acknowledges members of the European Spreadsheet Risks Interest Group discussion group, whose thoughtful comments and criticism helped sharpen his thinking through multiple drafts of this paper. Phil lives in Saint Louis, Missouri, USA, and is available for consultation in matters of spreadsheet style; he can be reached at xlPhil@gmail.com.